# Wavelength-Controlled Photocurrent Polarity Switching in BP-MoS$_2$ Heterostructure

Himani Jawa,[†] Sayantan Ghosh,[†] Abin Varghese,[†,‡] Srilagna Sahoo,[†] and Saurabh Lodha[*,†]

[†]*Department of Electrical Engineering, IIT Bombay, Mumbai, Maharashtra 400076*
[‡]*Department of Materials Science and Engineering, Monash University, Clayton, Victoria, Australia 3800*

E-mail: slodha@ee.iitb.ac.in

**Abstract**

Layered two-dimensional van der Waals (vdW) semiconductors and their heterostructures have been shown to exhibit positive photoconductance (PPC) in many studies. A few recent reports have demonstrated negative photoconductance (NPC) as well that can enable broadband photodetection besides multi-level optoelectronic logic and memory. Controllable and reversible switching between PPC and NPC is a key requirement for these applications. This report demonstrates visible-to-near infrared wavelength-driven NPC and PPC, along with reversible switching between the two, in an air stable, high mobility, broadband black phosphorus (BP) field effect transistor (FET) covered with a few layer MoS$_2$ flake. The crossover switching wavelength can be tuned by varying the MoS$_2$ bandgap through its flake thickness and the NPC and PPC photoresponsivities can be modulated using electrostatic gating as well as laser power. Recombination-driven NPC and PPC allows for reversible switching at reasonable time scales of a few seconds. Further, gate voltage-dependent negative persistent photoconductance enables synaptic behavior that is well-suited for optosynaptic applications.

# Introduction

Heterostructures based on layered van der Waals (vdW) materials have recently gained significant interest owing to the realization of a variety of optical[1–8] and electrical[9,10] phenomena. Stacking different 2D materials offers various advantages such as a wide optical detection range, enhanced photoresponse due to ease of carrier separation, clean interfaces due to the absence of dangling bonds and defects resulting from lattice mismatch. This makes them an excellent choice for photodetection applications. Generally, the device architecture and choice of 2D materials in these heterostructures have been shown to enhance the positive photoconductance[1,11–14] (PPC) for application areas such as photodetectors[15] or optical memories.[16,17] However, some recent studies have also shown the possibility of realizing negative photoconductance[18–23] (NPC) in 2D material-based devices.

The observation of negative photoconductance along with PPC in photodetectors can enable broadband photoresponse with enhanced spectral resolution[24] due to their ability to segregate different wavelengths with distinct conductance states. Furthermore, the availability of additional states (PPC and NPC) expands the application scope of 2D materials to multi-bit logic and memory devices. However, the ability to switch reversibly and controllably between NPC and PPC is a critical and unexplored requirement for the realization of logic, memory and optical communication technologies.[18] In addition, NPC, as observed in 2D materials-based devices,[25,26] could be related to trap centers,[27] interaction with adsorbates,[28] the bolometric effect[29] or dependent on gate bias.[20] The first three mechanisms increase carrier scattering in the device resulting in mobility degradation and negative photocurrent. Interface/adsorbate/bulk trap-related mechanisms are difficult to control since it is not trivial to engineer trap densities, distributions and time constants, whereas gate voltage-dependent NPC and PPC adds to the overall power consumption. Recently NPC has been reported in a $WS_2$/rGO hybrid structure under infrared light as a result of recombination between photogenerated electrons in $WS_2$ with holes in rGO.[19] NPC has also been reported in an ReS$_2$/hBN/MoS$_2$ heterostructure floating gate device with excellent memory properties.[18] These recent reports suggest new avenues for realizing NPC and controlled NPC-PPC switching based on heterostructures using 2D materials.

In this work, we demonstrate wavelength-driven (visible-to-near infrared) NPC and PPC, along with reversible switching between the two, in an air-stable BP FET covered with a few layer MoS$_2$ flake. BP is chosen for its high hole mobility and broad spectral response (visible as well as NIR) and MoS$_2$ is chosen for its air-stability, optical bandgap tunability and spectral response in the visible range. The NPC is a consequence of light absorption in the MoS$_2$ flake, which results in photocarrier generation. Photogenerated holes are captured by intrinsic traps in MoS$_2$ whereas the photogenerated electrons diffuse towards the BP/MoS$_2$ heterointerface to recombine with holes in the BP channel leading to a reduction in the net device current under illumination. As the wavelength is increased beyond the absorption edge of the MoS$_2$ flake, the phototransistor exhibits PPC solely based on the positive photoresponse of the BP flake. The critical wavelength for the crossover between NPC and PPC can be tuned through an appropriate choice of the MoS$_2$ flake thickness that determines its bandgap and absorption edge. The wavelength-driven NPC-PPC switching is reversible over multiple wavelength cycles, with rise and fall times of a few seconds, and the negative and positive photoresponsivities in the visible and NIR regimes can be tuned using electrostatic gating or laser power. Hence, a broad and tunable spectral response with distinct NPC and PPC regimes, that can be reversibly switched between using wavelength cycling, has been realized in a BP/MoS$_2$ heterostructure phototransistor. The presence of gate voltage-dependent negative persistent photoconductance makes this device promising for optosynaptic applications as well.

# Results and discussion

## Device Fabrication and Electrical Characterisation

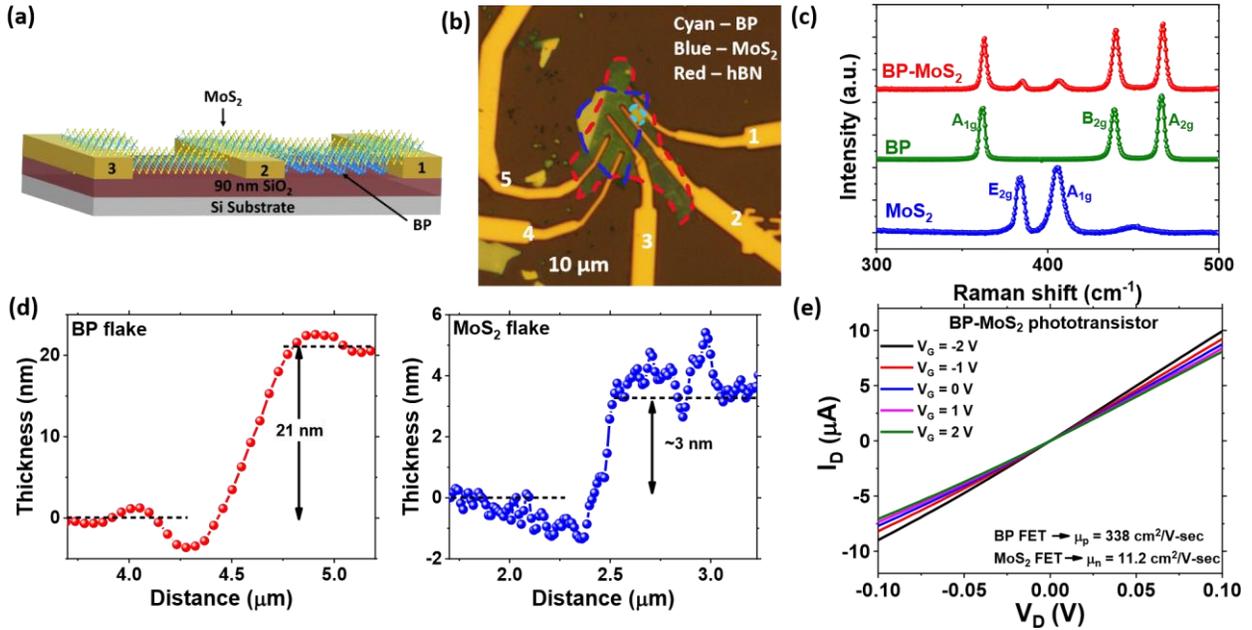

Figure 1: **Device schematic, fabrication and characterization.** (a) Schematic of BP/MoS$_2$ phototransistor (b) Optical microscope image showing BP/MoS$_2$ phototransistor with source/drain metal contacts sandwiched between BP and MoS$_2$ layers. hBN flake on the top precludes any ambient effects. (c) Raman spectra acquired for the BP/MoS$_2$ phototransistor (d) AFM scan for BP and MoS$_2$ flakes (e) Output characteristics of the phototransistor showing ohmic contacts and low gate modulation.

BP flakes were mechanically exfoliated on a 90 nm SiO$_2$/Si substrate using scotch tape. Source/drain contacts were then formed on the few layer BP flake using electron beam lithography (EBL) and metal deposition (Cr/Au ~ 5/40 nm). An additional contact was fabricated adjacent to the source/drain contacts of the BP FET. Next, MoS$_2$ flakes were mechanically exfoliated onto a polydimethylsiloxane (PDMS) stamp and large, thin flakes were transferred on to the FET to ensure complete coverage of the BP flake and the extra contact. Further, contacts were formed on the MoS$_2$ flake using EBL and metal deposition (Cr/Au ~ 5/40 nm). Table S1 shows a set of 10 such devices with their respective BP and MoS$_2$ flake thicknesses. Figure 1a shows a schematic of the BP/MoS$_2$ phototransistor (device 1). An optical microscope image of the fabricated device is shown in Figure 1b with the phototransistor covered with an hBN flake (~ 10 nm) to preclude any trapping effects from adsorbates on the electrical and optical characterisation of the phototransistor. The contacts, labelled in the figure are as follows: contacts 1 and 2 sandwiched between BP and MoS$_2$ flakes for BP/MoS$_2$ phototransistor, contacts 2 and 3 being formed below the MoS$_2$ flake in order to quantitatively segregate its contribution in the phototransistor, and contacts 4 and 5 formed on top of the MoS$_2$ flake to understand its individual optical behavior. To check the crystallinity of the flakes after fabrication, Raman scans for individual materials as well as the overlap region were done. As shown in Figure 1c, Raman peaks at 384 and 405.56 cm$^{-1}$ correspond to E$_{2g}$ (in-plane vibration) and A$_{1g}$ (out-of-plane vibration) modes for MoS$_2$ whereas peaks at 363.1, 439.7 and 467.5 cm$^{-1}$ indicate the A$_{1g}$, B$_{2g}$ and A$_{2g}$ peaks for the BP flake. The characteristic peaks for both materials being distinctly present in the overlap region confirms the crystalline nature of the exfoliated flakes. Further, thickness of BP and MoS$_2$ flakes obtained using atomic force microscopy (AFM) are shown in Figure 1d.

Individual field effect transistor performances were studied for the BP and MoS$_2$ FETs. Their transfer and output characteristics are shown in Figure S1 of the Supporting Information. The BP FET, with a hole mobility of 338 cm$^2$/V-sec (calculated at maximum transconductance) shows slight gate modulation in transfer characteristics (I$_D$-V$_G$), shown in Figure S1a, with a high positive threshold voltage (> 5 V). On the other hand, the MoS$_2$ FET, measured between contacts 4 and 5, shows good gate tunability of I$_D$ (shown in Figure S1c) with an electron mobility of ~11 cm$^2$/V-sec. The output characteristics of the MoS$_2$ FET with contacts below (contacts 2 and 3) and above the flake (contacts 4 and 5) highlight their ohmic behavior with the currents being slightly higher for contacts 4 and 5 made above the flake, as shown in Figure S1b and S1d. The output (I$_D$-V$_D$) characteristics of the phototransistor, as shown in Figure 1e, indicate a major contribution to the current from the BP flake (~ $\mu$As) as compared to a few nAs from the MoS$_2$ flake with low gate modulation.

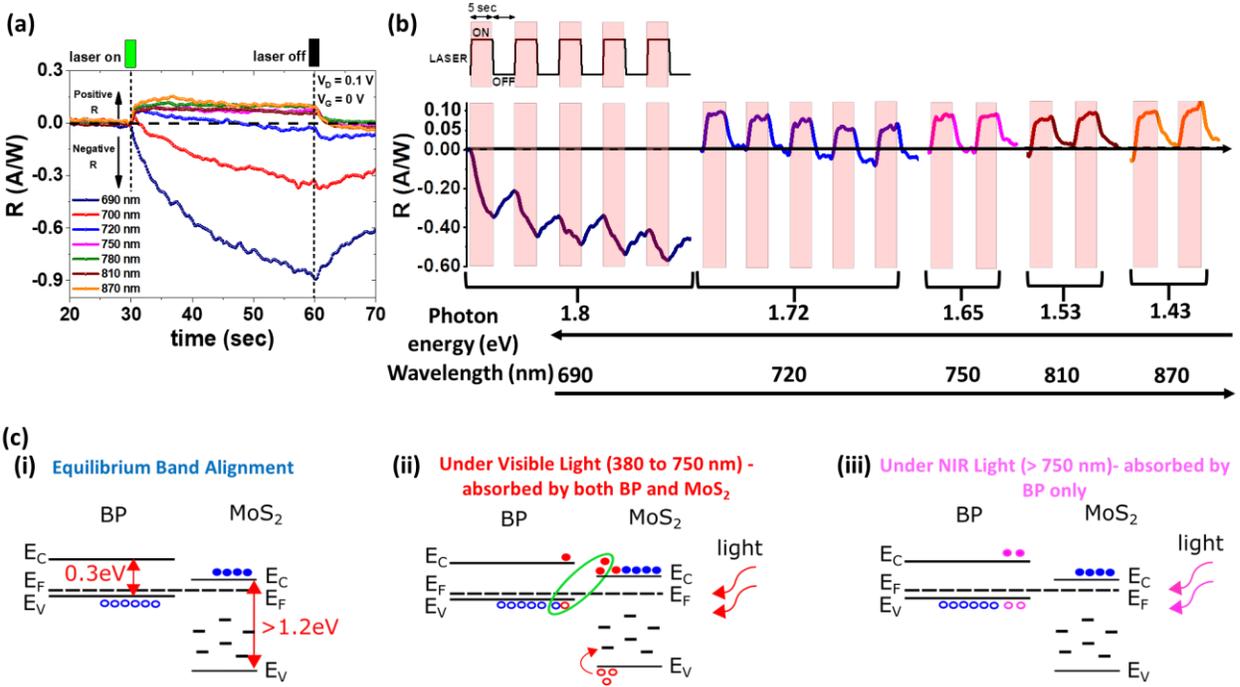

Figure 2: **Wavelength-dependent photoconductance and related band diagrams.** (a) Responsivity versus time for the BP/MoS$_2$ phototransistor showing negative responsivity for $\lambda < 750$ nm and positive responsivity for $\lambda > 750$ nm at $V_G = 0$ V and $V_D = 0.1$ V (b) Photoswitching with on and off time being 5 sec at $V_G = 0$ V and $V_D = 0.1$ V for different wavelengths. (c) BP-MoS$_2$ energy band alignment at $V_G = 0$ V for (i) equilibrium conditions, (ii) under visible light, where both BP and MoS$_2$ absorb light resulting in interlayer recombination and hence, negative photocurrent, and (iii) for NIR illumination, which is absorbed only by the BP flake and hence, positive photocurrent is observed.

## Optoelectronic Characterisation of the BP/MoS$_2$ Phototransistor

Photoresponse of the BP/MoS$_2$ heterostructure was obtained under 690-900 nm laser illumination. The light was incident on the phototransistor for 30 seconds and then turned off. The photoresponsivity (R) was calculated using the equation: R = $I_{ph}/P_{in}$, where $I_{ph}$ = $I_{light}$ - $I_{dark}$ with $I_{light}$ and $I_{dark}$ being the currents with and without illumination and $P_{in}$ is the incident optical power. As shown in Figure 2a, responsivity is negative for wavelengths ($\lambda$) upto 720 nm and positive beyond 750 nm for gate ($V_G$) and drain ($V_D$) voltages of 0 and 0.1 V respectively. The negative R can be explained from the band alignment of the BP and MoS$_2$ flakes in the transverse direction (along gate-SiO$_2$-BP-MoS$_2$). Under equilibrium, BP/MoS$_2$ forms a type-II heterostructure with a band alignment as shown in Figure 2c(i). Both BP and MoS$_2$ absorb visible light (λ < 750 nm) illuminated on the overlap region, with MoS$_2$ absorbing more than BP, thereby leading to generation of electron-hole pairs. The photogenerated holes in MoS$_2$ get captured by the shallow intrinsic traps distributed near its valence band[30] giving rise to excess electron concentration. These excess electrons diffuse towards BP due to the concentration gradient set up by a lower concentration in BP, leading to interlayer recombination with the high hole concentration of BP (Fig. 2c(ii)). This results in a reduction in free hole concentration in BP giving rise to a negative photocurrent or responsivity (as seen in Figure 2a), since BP is the main photoconducting layer in this BP-MoS$_2$ stack. As the wavelength is increased beyond 750 nm, a positive I$_{ph}$ is observed which can be attributed to (i) negligible optical response from the thin MoS$_2$ flake limited by its bandgap and (ii) photogenerated carriers in the smaller bandgap (0.3 eV) BP flake. The optical response of the MoS$_2$ FET (contacts 4 and 5) to 690 and 720 nm incident light is shown in Figure S2a of Supporting Information. MoS$_2$ does not respond to illumination beyond 720 nm due to its bandgap and therefore, the crossover wavelength λ$_t$, defined as the wavelength at which the I$_{ph}$ transitions from negative to positive values, is limited by the bandgap of MoS$_2$. Figure 2b and S3a of Supporting Information show the transition from negative to positive responsivity and photocurrent when the phototransistor was illuminated under different wavelengths with the laser being switched on and off every 5 seconds for each λ.

**Gate Voltage and Laser Power Dependent Photocurrent Modulation**: The interlayer recombination occurring at the BP/MoS$_2$ interface under visible light, as discussed above, depends on (i) the electron concentration of MoS$_2$ (n$_M$) and (ii) hole concentration of BP (p$_{BP}$). These carrier concentrations are modulated by the applied gate bias electrostatically and optically and light intensity respectively. Also, the hole concentration in BP is affected by the trapping and detrapping at the BP/SiO$_2$ interface which varies with applied gate bias.[31] This results in a dual effect of V$_G$ on the carrier concentrations and hence, on the interlayer carrier recombination. Under illumination for varying $V_G$, the $I_{ph}$ (dependent on carrier recombination) will be maximum where the effective electron and holes concentrations in MoS$_2$ and BP respectively will be nearly equal (as shown in Figure S4). The $I_{ph}$ will gradually reduce with $V_G$ on both sides of the maximum point due to reduction in either holes or electrons at positive and negative $V_G$ respectively, since the recombination is limited by the minority carrier concentration. This results in a U-shaped photocurrent plot ($I_{ph}$ vs $V_G$ at P$_{in1}$ in Figure S4b).

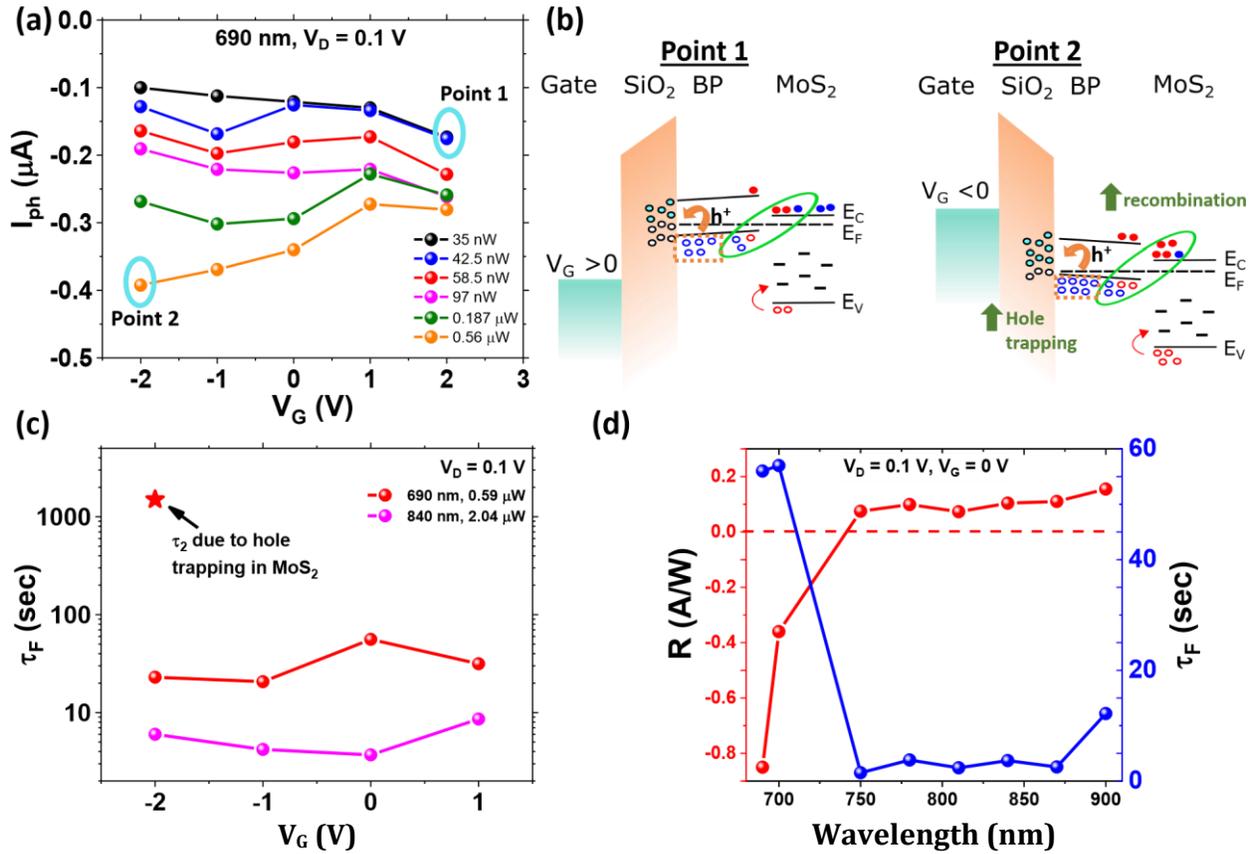

Figure 3: **Gate voltage, laser power and wavelength dependence of temporal response and photocurrent.** (a) Variation of photocurrent as a function of gate voltage under 690 nm illumination (b) Energy band diagrams at points 1 and 2 of figure (a) (c) Fall time ($\tau_F$) vs $V_G$ for 690 and 840 nm illumination indicates faster operation under NIR (d) Responsivity (R) and $\tau_F$ as a function of wavelength. Negative responsivity and higher fall times correspond to the optical sensing of MoS$_2$ in the phototransistor.

Figure 3a shows the variation in photocurrent with gate voltage for varying incident power of 690 nm illumination. The photocurrent magnitude shows a slight increase with increasing gate voltage at low laser powers, a U-shaped variation at intermediate powers (e.g. at 0.187 $\mu$W, similar to the theoretical expectation illustrated in Figure S4b) and a more pronounced change (decrease) in its value at higher optical powers. At lower optical power, most of the electron-hole pair generation takes place in MoS$_2$. Given the trapping of photogenerated holes in MoS$_2$ and a high hole concentration in BP, the interlayer recombination is limited by the MoS$_2$ electron concentration ($n_M < p_{BP}$). The electron concentration in MoS$_2$ increases with increase in gate voltage (as the device transitions from subthreshold to ON state, as shown in Figure S1c of Supporting Information). Hence, a slightly higher interlayer recombination at positive $V_G$ leads to slightly higher $I_{ph}$ as shown by the band diagram for point 1 in Figure 3b. At lower optical powers, we observe only the left (increasing) $I_{ph}$ branch of the U-shaped curve at $P_{in1}$ in Figure S4b. As the optical intensity is increased, $n_M$ increases due to an increase in the photogenerated carriers in MoS$_2$. This leads to interlayer recombination being limited by $p_{BP}$, which gets affected by the trapping and detrapping of holes at the BP/SiO$_2$ interface.[31] As $V_G$ is reduced from positive to negative values, the traps at the BP/SiO$_2$ interface above the Fermi level ($E_F$) of BP get filled with holes, going from neutral to positively charged. This results in the reduction of free hole carrier concentration in BP as well as of the effective negative gate bias, similar to the left-shift in the U-shaped curve at higher optical power $P_{in2}$ of Figure S4b. In order to compensate for this reduction in hole concentration, a more negative $V_G$ is required and hence, maximum photocurrent is shifted towards negative $V_G$, as shown at point 2 in Figure 3b. In the case of higher optical powers, we observe only the right (decreasing) branch of the U-shaped $I_{ph}$ curve at $P_{in2}$ in Figure S4b.

When the phototransistor is illuminated with 780 nm laser (NIR), a positive photocurrent is observed, as shown in Figures S3b and S3c of Supporting Information for varying optical power and gate voltage respectively. This positive $I_{ph}$ increases as the input incident power increases due to an increase in the photogenerated carrier concentrations in the BP flake, as shown in Figure S3b of Supporting Information. The photocurrent also increases as the gate voltage is increased from negative to positive values. This can be explained by increased concentration of photogenerated free carriers at positive $V_G$ as compared to negative $V_G$ with the maximum photocurrent occurring at maximum transconductance value.[32] This is due to dominant photogating effect at the maximum transconductance operating point. Since the BP FET is in the accumulation region for the given $V_G$ range, the $I_{ph}$ increases with increase in $V_G$.

**Temporal Photoresponse**: Next, temporal measurements were carried out in order to understand the variation in the speed of the device for negative and positive photocurrent regimes. As shown in Figure 3c, the fall time ($\tau_F$) remains nearly constant over gate voltage for both 690 (visible) and 840 nm (NIR) illumination. However, we observe that (i) $\tau_F$ is lower in the positive $I_{ph}$ regime i.e. under 840 nm illumination compared to 690 nm and (ii) there is a second time constant at $V_G$ = -2 V (~ 1.5 x 10$^3$ sec) which arises from hole trapping in the MoS$_2$ flake (discussed later in this paper). Under 690 nm illumination, free carriers in both flakes (p$_{BP}$ and n$_M$) are responsible for photoconduction and negative $I_{ph}$. The slow response of the MoS$_2$ flake (as can be observed from the slow recovery of the MoS$_2$ FET after the laser is switched off, shown in Figure S2a of Supporting Information) due to long detrapping time of the photogenerated holes results in increased lifetime of the photogenerated electrons thereby leading to a large $\tau_F$.[33] However, under 840 nm illumination, the positive $I_{ph}$ arises only from the photogenerated carriers of BP thereby resulting in relatively faster optical operation or lower fall time. This can also be observed in Figure 2b where the photocurrent $I_{ph}$ (or photoresponsivity, R) increases in magnitude with each optical cycle under visible light as it is limited by the time constant of the carriers in the MoS$_2$ flake, however a steady on and off cycles are observed for NIR illumination due to fast hole detrapping time constant in the BP flake. Figure 3d shows variation in $\tau_F$ and R with wavelength. For $\lambda$ < 750 nm, $\tau_F$ is higher (~ 55 sec) due to MoS$_2$-limited dark current recovery. For $\lambda$ beyond 750 nm, lower fall times are observed due to higher mobility in the BP flake and faster detrapping time for trapped photocarriers from shallow traps in BP bulk.[32]

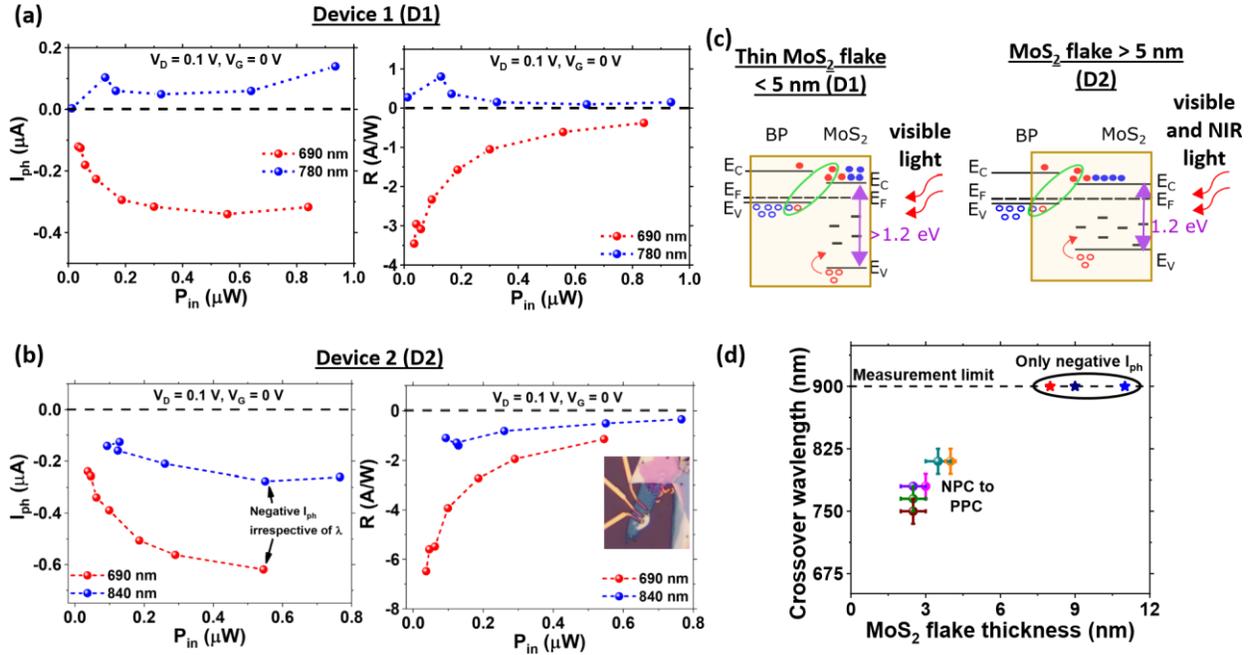

Figure 4: **MoS$_2$ thickness dependent NPC and PPC crossover wavelength**- Photocurrent and responsivity as a function of input laser power for (a) device 1 (MoS$_2$ thickness of ~3 nm) and (b) device 2 (MoS$_2$ thickness of 11 nm) under 690 and 840 nm illumination. Negative photocurrent in visible and NIR corroborates absorption of light by both MoS$_2$ and BP flakes. (c) Band alignment in BP and MoS$_2$ with varying thickness of MoS$_2$ flake. (d) MoS$_2$ thickness vs crossover wavelength, $\lambda_t$ (wavelength at which the photocurrent transitions from negative to positive values).

## MoS$_2$ Thickness Dependent Negative Photocurrent

The negative photocurrent depends on the free carrier concentrations of BP and MoS$_2$ flakes that can be modulated in magnitude (photoresponsivity) and spectral range by the gate voltage and the optical absorption range of the MoS$_2$ flake respectively. Given that light absorption by the MoS$_2$ flake results in generation of photocarriers that diffuse towards BP resulting in interlayer recombination and hence, negative $I_{ph}$, the optical spectral response of MoS$_2$ determines the crossover wavelength from NPC to PPC for the phototransistor. The spectral response of MoS$_2$ is limited by its bandgap which varies with the thickness of the flake and therefore, the crossover wavelength increases for the first few nanometers of the MoS$_2$ flake as the bandgap decreases with increasing thickness, till the bandgap becomes constant (~ 1.2 eV) for a few layer MoS$_2$ flake. Figure 4a-c shows the optical response and corresponding band diagrams for two phototransistors with ~3 nm and 11 nm thick MoS$_2$ flakes. Device 1 (flake thickness ~3 nm) shows negative (positive) I$_{ph}$ and R for 690 nm (780 nm) as its $\lambda_t$ is approximately 750 nm (~ 1.65 eV) which is close to the bandgap of the MoS$_2$ flake. On the other hand, the 11 nm thick MoS$_2$ flake in device 2 has a bandgap of ~1.2 eV. Hence, it exhibits a negative optical response to both 690 and 840 nm illumination since its $\lambda_t$ is greater than 1000 nm (which is difficult to demonstrate given the limitation of our measurement setup). As the flake thickness of MoS$_2$ increases further, the diffusion of electrons from the top surface (high optical absorption region) of the MoS$_2$ flake towards BP becomes less favorable. This leads to negligible interlayer recombination resulting in a direct flow of electrons from conduction band of MoS$_2$ to the contacts. This leads to a positive I$_{ph}$ for visible and NIR illumination, as shown in Figure S5 (device 3). It should be noted that a few layer BP flake is selected for these phototransistors with a reasonable I$_{on}$/I$_{off}$ and a fixed bandgap of ~ 0.3 eV. Figure 4d shows the variation of crossover wavelength with MoS$_2$ thickness for multiple devices indicating that as the MoS$_2$ flake thickness increases, $\lambda_t$ increases till it saturates for few layer flakes.

## Applications

The BP/MoS$_2$ device architecture gives the advantage of switching the photocurrent from negative to positive values within a time scale of a few seconds as the wavelength switches from 700 to 750 nm, as shown in Figure 5a. Photocurrent switching characteristics of two additional devices are shown in Figure S6. This transition can be used in a wavelength dependent multi-bit coding scheme as it helps in differentiating change in current due to different wavelengths of light thereby increasing the accuracy of light recognition. Further, this device also demonstrates negative persistent photoconductance, where the photocurrent reaches a maximum negative value under illumination and when the light is withdrawn, it holds on to a negative current value (lower than the maximum but not the same as the dark current) for a long period of time. Figure 5b shows negative persistent photoconductance of device 1 for $V_G$ = -2 V with the normalised photocurrent Y (shown in Fig 5b inset) following a double exponential function: $Y = A_1 exp(-t/\tau_1) + A_2 exp(-t/\tau_2)$.[34] Here t is time, $A_1$ and $A_2$ are constants and $\tau_1$ and $\tau_2$ are fast and slow decay time constants respectively. The faster time constant, $\tau_1$ = 23 sec symbolises rapid relaxation of the device's photocurrent in response to light withdrawal, whereas the slower time constant $\tau_2$ = 1.5 x $10^3$ sec arises from hole trapping in $MoS_2$ on visible light illumination which results in slower decay in current. This behavior is similar to the long-term potentiation observed in synaptic devices. Similar behavior for another device (device 6) along with all its other optical measurements is shown in Figure S7.

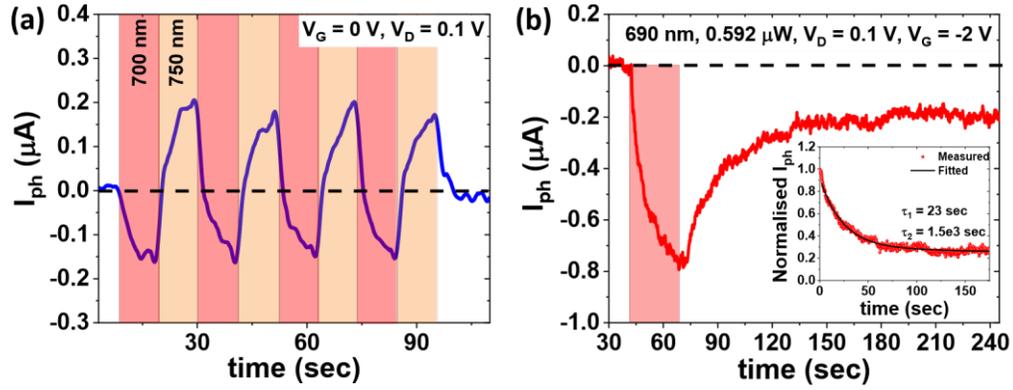

Figure 5: **Wavelength-based photoswitching and negative persistent photoconductance** (a) Wavelength-based photocurrent switching at $V_G$ = 0 V and $V_D$ = 0.1 V (b) Negative persistent photoconductance at $V_G$ = -2 V and $V_D$ = 0.1 V with the inset showing a double exponential-based fitting of the normalised photocurrent.

## Conclusions

This work demonstrates wavelength-dependent photoconductance polarity and reversible NPC-PPC switching in a BP/$MoS_2$ phototransistor. Table 1 compares the key optical characteristics of various architectures and devices using 2D materials that demonstrate negative photoconductance along with the BP/$MoS_2$ phototransistor reported in this work.

The NPC is a result of recombination between photogenerated electrons in MoS$_2$ with holes in the BP flake. The NPC-to-PPC crossover wavelength is tunable with the thickness of the MoS$_2$ flake. MoS$_2$ thickness modulates its bandgap and optical absorption range thereby resulting in NPC (PPC) for wavelengths within (beyond) the absorption range. The device offers dual fold application: first, different photoelectric response to different wavelengths, owing to the presence of both NPC and PPC in the same device, expands its application prospects to broadband photoelectric sensing and second, the presence of negative persistent photoconductance at negative gate voltage offers a promising avenue for opto-controlled synaptic applications.

## Experimental Procedure

The BP transistors were fabricated using a degenerately doped p-type Si substrate with 280 nm thermally grown SiO$_2$. BP flakes were mechanically exfoliated using scotch tape and transferred onto the substrate. The flakes were then identified using an optical microscope and source/drain contacts (along with an additional adjacent contact) were patterned by electron beam lithography (Raith 150-Two) using poly(methyl methacrylate) resist. Source/drain metal (Cr/Au -5/40 nm) was deposited using sputtering followed by lift-off. Further, MoS$_2$ flakes were mechanically exfoliated from the molybdenite crystal using polydimethylsiloxane (PDMS) and were transferred on the fabricated BP FET using a pick-and-transfer process using a micromanipulator setup. The contacts on MoS$_2$ were fabricated using e-beam lithography and metal deposition. The final device image was taken using a Olympus BX-63 microscope and the SEM imaging was done using Raith 150-Two. ULVAC-PHI/PHI5000 Versa ProbeII focus X-ray photoelectron spectrometer was used for XPS measurements and Horiba HR 800 Raman spectroscopy system with a 532 nm laser was used for Raman imaging. Before optical characterization, the device was placed on a PCB with large gold contact pads. The device contact pads were wire bonded using gold wire to the large PCB contact pads. The optoelectronic measurements were done in ambient conditions under a BX-63 Olympus microscope using a Keysight B1500A semiconductor device analyzer using an NKT laser with a wavelength range of 690 to 900 nm. Switching of laser power was done using input of square pulses of 10 V peak-to-peak to the laser power controller unit from an Agilent 33220A function generator. The laser was incident on the device through the objective lens of the BX-63 Olympus microscope.

Table 1: Comparison of this work with previous studies demonstrating negative photoconductance.

| Ref | Material | Spectral response (nm) | NPC and PPC | NPC ↔ PPC modulator | Mechanism | NPC-to-PPC switching | | Application |
|---|---|---|---|---|---|---|---|---|
| | | | | | | Modulator | Reversible | |
| [29] | BP | 830 | only NPC | NIR | bolometric effect | -- | -- | -- |
| [35] | MoS$_2$ | NIR | only NPC | Monolayer MoS$_2$ | trion formation | -- | -- | -- |
| [36] | MoS$_2$ | 454, 519, 625, 980 and 1550 | both | NIR | bolometric effect | -- | -- | -- |
| [18] | ReS$_2$/hBN/MoS$_2$ | 520, 637, 830 and 1310 | both | $V_G$ and $\lambda$ | charge trapping | -- | -- | Photoelectronic memory |
| [19] | WS$_2$/RGO | 808 | only NPC | NIR | recombination | -- | -- | -- |
| [20] | BP/SnS$_{0.5}$Se$_{1.5}$ | 365 and 894.6 | both | $V_G$ | charge trapping | -- | -- | -- |
| [21] | Gr/BP | 655, 785 and 980 | both | $V_G$ | electron trapping | -- | -- | -- |
| [22] | MoSe$_2$/Gr | 450 to 1000 | both | $V_G$ | charge transfer | -- | -- | -- |
| [23] | Gr/MoS$_2$ | 635 | both | $V_G$ | charge transfer | -- | -- | -- |
| [37] | MoTe$_2$/Gr | 975 | both | laser power | charge transfer | laser power | Yes | -- |
| **This work** | **BP/MoS$_2$** | **690 to 900** | **both** | **$\lambda$ ~ MoS$_2$ bandgap** | **recombination** | **$\lambda$** | **Yes** | **Photoelectronic memory** |

## Acknowledgement

The authors acknowledge Indian Institute of Technology Bombay Nanofabrication Facility (IITBNF) for the device fabrication and characterization. The authors thank Kartikey Thakar for wire bonding the samples. H.J. acknowledges Visvesvaraya PhD Scheme from Ministry of Electronics and Information Technology (Meity), Govt. of India. A.V. thanks IITB-Monash Research Academy for the doctoral fellowship and S.L. acknowledges the Department of Science and Technology (DST), Govt. of India through its SwarnaJayanti fellowship scheme (Grant number - DST/SJF/ETA-01/2016-17) for funding support.

## Supporting Information Available

The supporting information consists of table with data set of 10 devices, transfer and output characteristics of BP and $MoS_2$ FETs (device 1), optical characteristics of $MoS_2$ FET (device 1), optical characteristics of multiple phototransistors (device 1, device 3, device 4, device 5 and device 6).

Supporting Information

# Wavelength-Controlled Photocurrent Polarity Switching in BP-MoS$_2$ Heterostructure

Himani Jawa,[†] Sayantan Ghosh,[†] Abin Varghese,[†,‡] Srilagna Sahoo,[†] and Saurabh Lodha[*,†]

[†]*Department of Electrical Engineering, IIT Bombay, Mumbai, Maharashtra 400076*

[‡]*Department of Materials Science and Engineering, Monash University, Clayton, Victoria,*

*Australia 3800*

E-mail: slodha@ee.iitb.ac.in

# Details of the Fabricated BP-MoS$_2$ Phototransistors

Table S1: MoS$_2$ and BP thicknesses for the various phototransistors fabricated in this study.

| Device | MoS$_2$ thickness (nm) | BP thickness (nm) | NPC and PPC |
|--------|------------------------|-------------------|-------------|
| 1      | ~3                     | 21                | both        |
| 2      | 11                     | few layer         | only NPC    |
| 3      | 14                     | 8                 | only PPC    |
| 4      | ~2.5                   | ~9                | both        |
| 5      | ~3                     | ~33               | both        |
| 6      | ~3.5                   | 15                | both        |
| 7      | 8                      | 3                 | only NPC    |
| 8      | 9                      | ~30               | only NPC    |
| 9      | ~2.5                   | few layer         | both        |
| 10     | 4                      | few layer         | both        |

# Transfer and Output Characteristics of BP and MoS$_2$ FETs

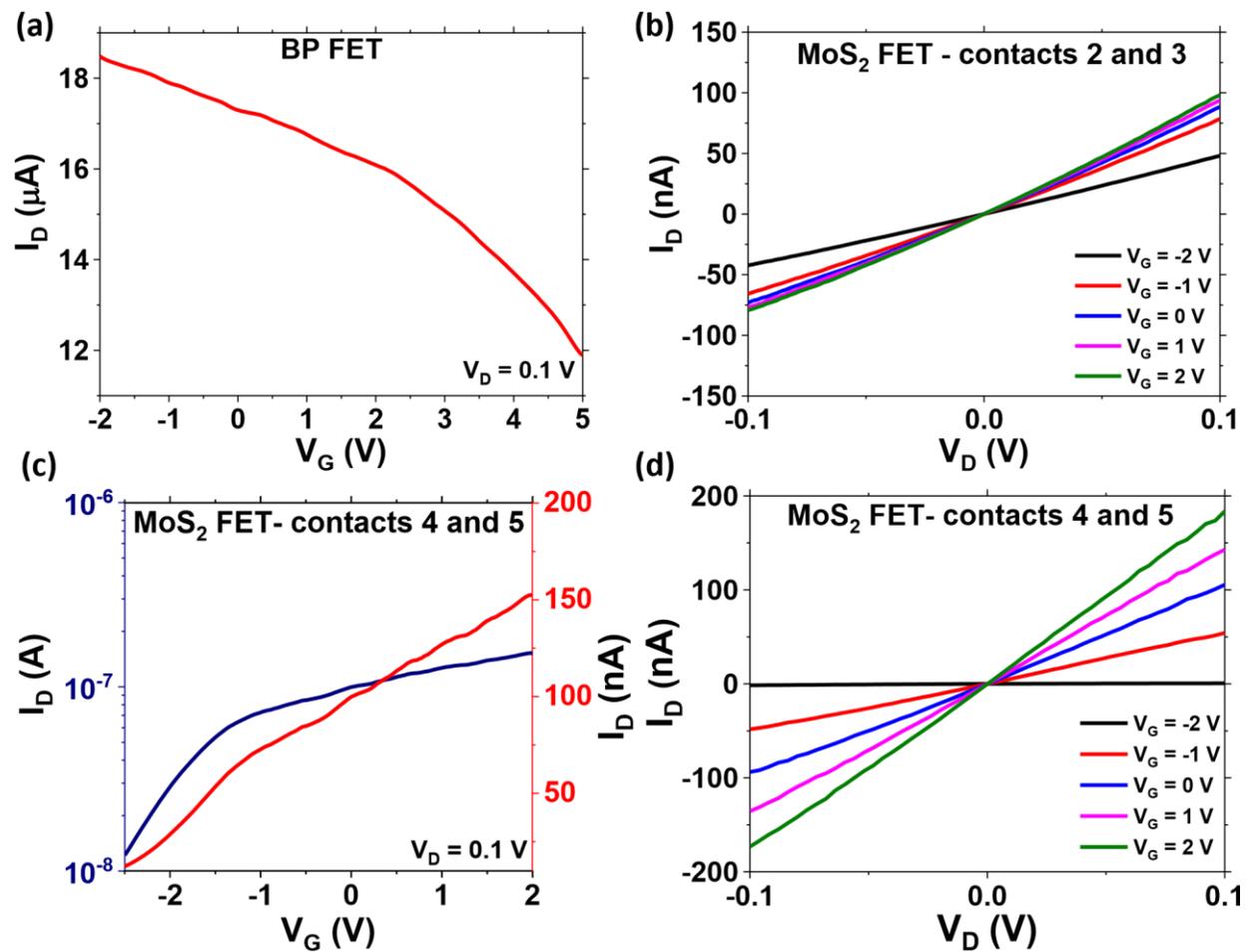

Figure S1: (a) Transfer characteristics of the BP FET (b) Output characteristics of the MoS$_2$ FET between contacts 2 and 3 (contacts are below the MoS$_2$ flake) (c) Transfer and (d) output characteristics of the MoS$_2$ FET between contacts 4 and 5 (contacts fabricated on top of the MoS$_2$ flake).

## Optical Characteristics of the MoS$_2$ FET

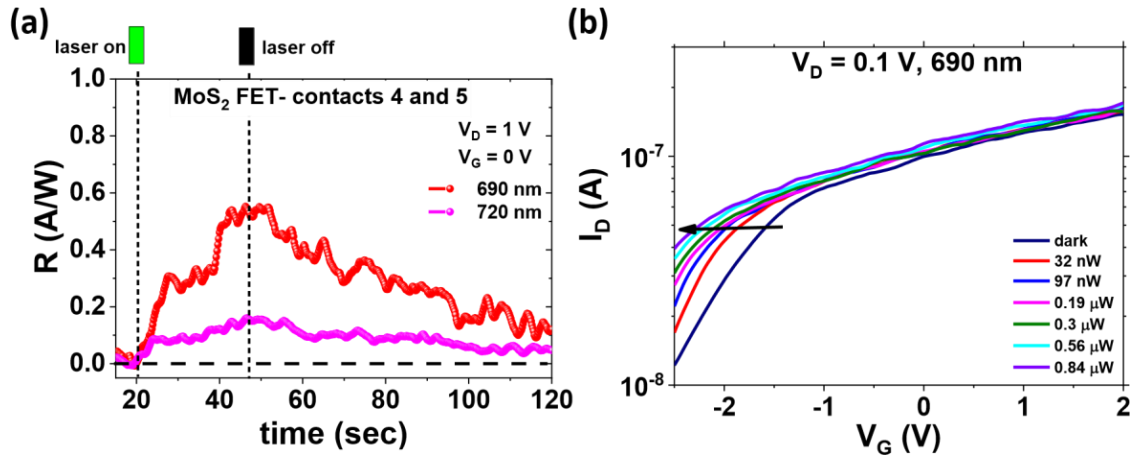

Figure S2: (a) Optical response of the MoS$_2$ FET under 690 and 720 nm illumination at $V_G$ = 0 V and $V_D$ = 1 V (b) Transfer characteristics of the MoS$_2$ FET for varying incident laser power of 690 nm at $V_D$ = 0.1 V showing photogating effect due to trapped photogenerated holes.

## Optical Characteristics of the BP-MoS$_2$ Phototransistor

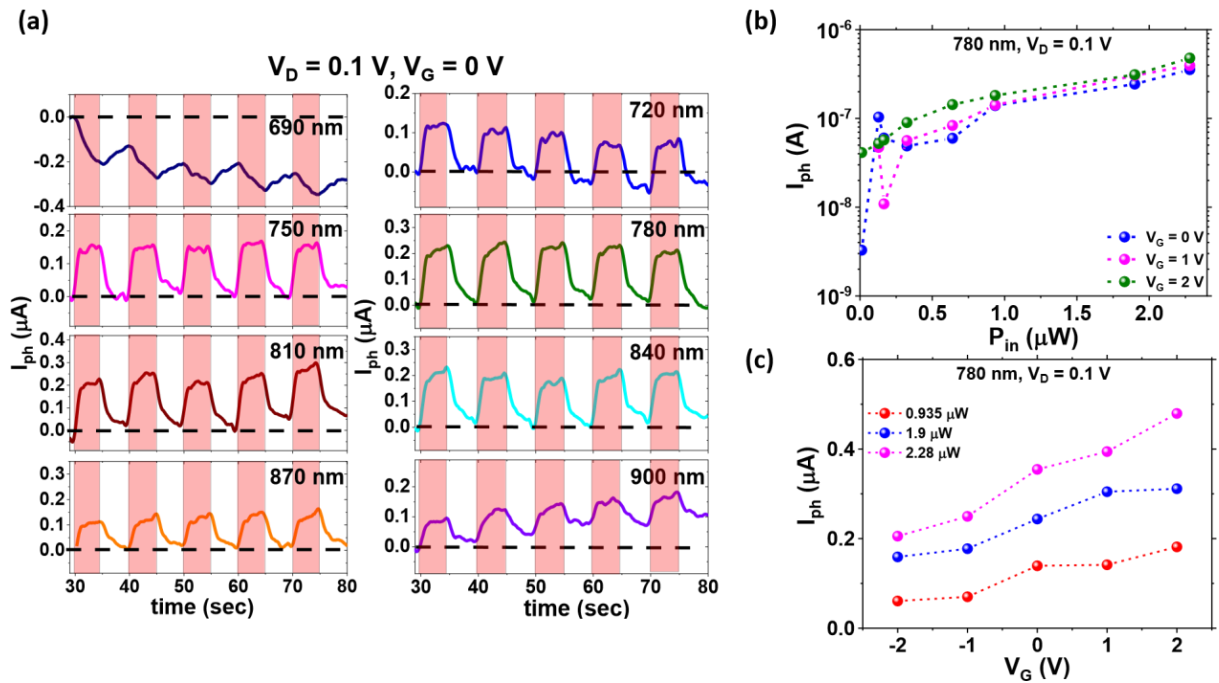

Figure S3: (a) Photoswitching of device 1 for wavelengths ranging from 690 to 900 nm at $V_G$ = 0 V and $V_D$ = 0.1 V (b) Photocurrent vs incident laser power and (c) Variation of photocurrent with gate voltage under 780 nm illumination.

# Model for Photocurrent Modulation with Gate Voltage and Laser Power

Under 690 nm illumination, $n_M$ (dotted blue curve) and $p_{BP}$ (dotted red curve), as shown in Figure S4a, are the total electron and hole concentrations at laser power $P_{in1}$ that can be modulated with $V_G$. Carrier recombination, which determines the $I_{ph}$, depends on both $n_M$ and $p_{BP}$ resulting in a U-shaped curve (green curve). As the laser power is increased to $P_{in2}$, under ideal conditions, $n_M \times p_{BP}$ increases, given the increase in the photogenerated carriers (shown by the black curve in Figure S4a). However, due to trapping at the BP/SiO$_2$ interface, the hole concentration is limited in BP (although photogenerated electrons are available in MoS$_2$). This results in the left shift of the $n_M \times p_{BP}$ plot and hence, the $I_{ph}$, with the maximum occurring at negative $V_G$ as shown by the pink curve.

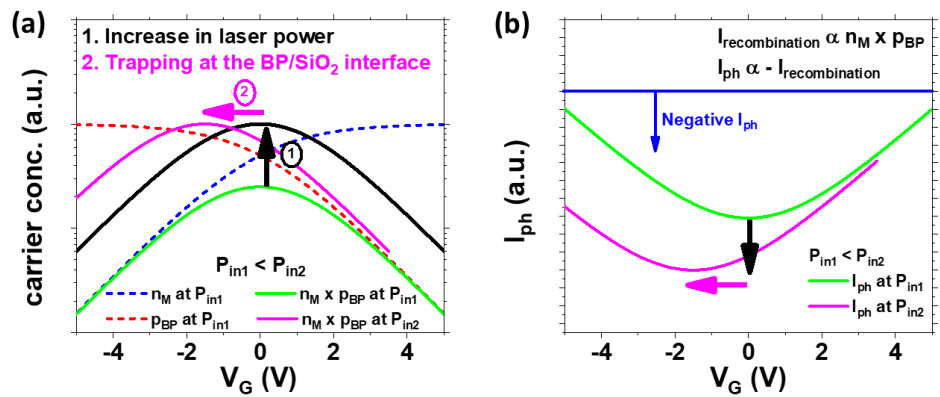

Figure S4: Model for (a) carrier concentration and (b) photocurrent variation in BP/MoS$_2$ phototransistor with gate voltage and optical power.

# Band Alignment and Optical Characteristics of Device 3

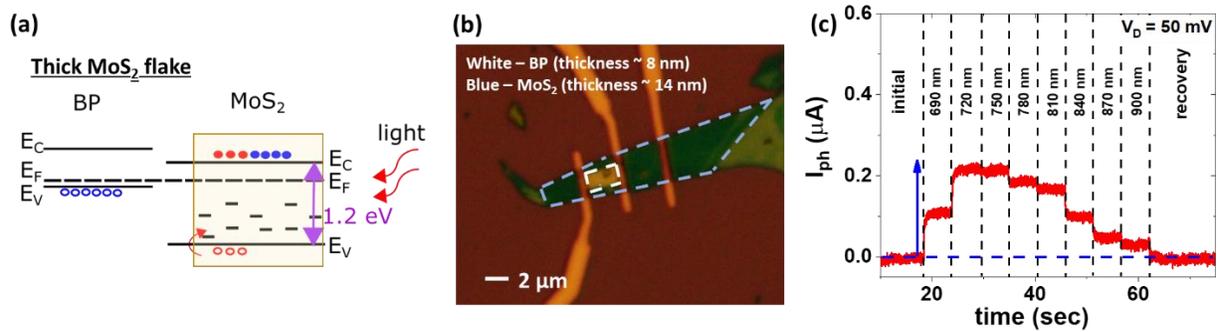

Figure S5: (a) Band alignment between BP and MoS2 for a thick MoS2 flake. (b) Optical image of device 3. (c) Photocurrent response of device 3 under different wavelengths showing a positive photocurrent irrespective of the wavelength.

# Photoswitching Characteristics of Device 4 and 5

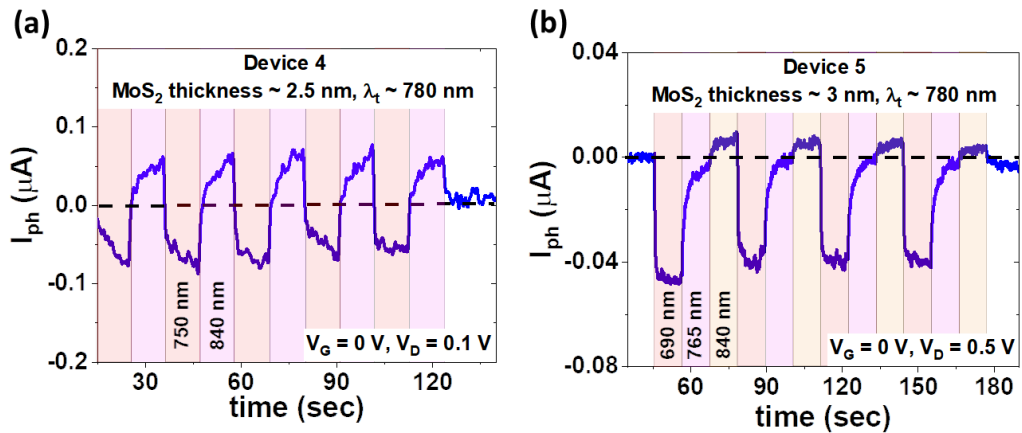

Figure S6: Wavelength-based photoswitching of (a) Device 4 and (b) Device 5.

## Optical Characteristics of Device 6

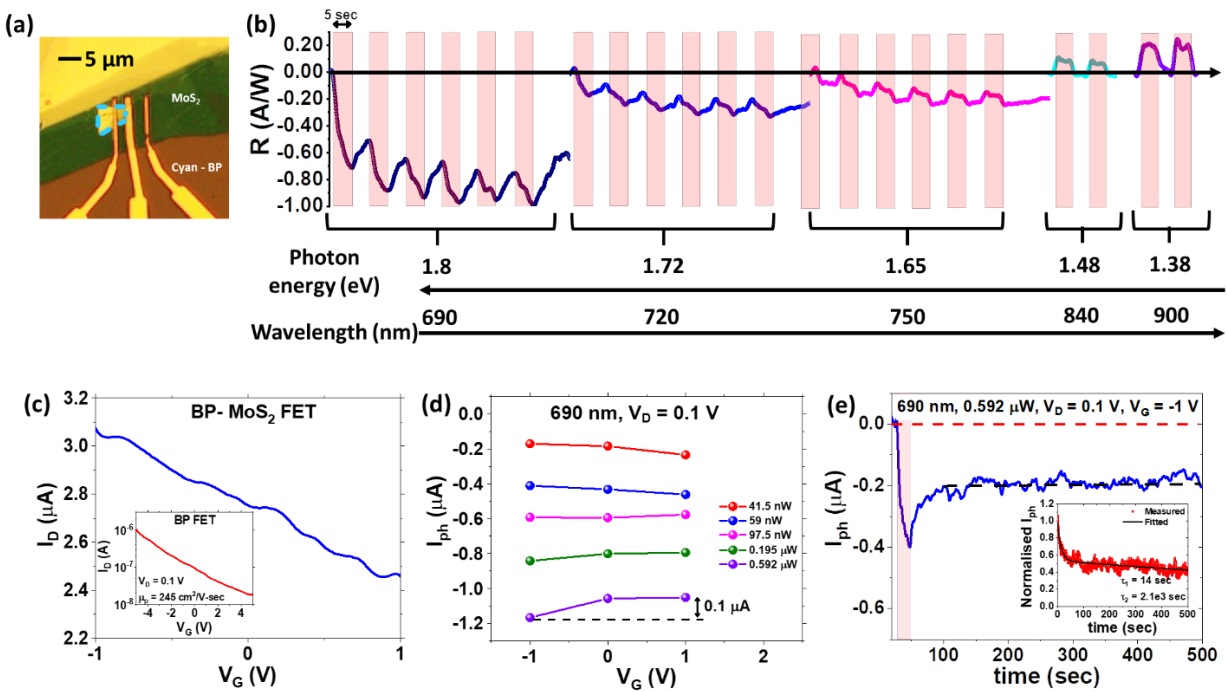

Figure S7: (a) Optical image of device 6 (b) Photoswitching with on and off time being 5 sec at $V_G$ = 0 V and $V_D$ = 0.1 V for different wavelengths. (c) Transfer characteristics of BP/MoS$_2$ device with the inset showing the transfer characteristics of the BP FET (d) Variation in photocurrent as a function of gate voltage under 690 nm illumination (e) Negative persistent photoconductance at $V_G$ = -1 V and $V_D$ = 0.1 V with the inset showing a double exponential-based fitting of the normalised photocurrent.